\documentclass{PoS}

\usepackage{amsmath}
\usepackage{amssymb}
\usepackage{color}
\usepackage{float}

\title{Rare FCNC radiative leptonic decays $B\to\gamma\ell^+\ell^-$}

\ShortTitle{FCNC radiative leptonic $B$-decays}

\author{\speaker{Dmitri Melikhov}$^{abc}$, Anastasiia Kozachuk$^{bd}$
and Nikolai Nikitin$^{bd}$\\
\llap{$^a$}
Institute for High Energy Physics, Austrian Academy of Sciences, Nikolsdorfergasse 18, A-1050 Vienna, Austria\\
\llap{$^b$}
D.~V.~Skobeltsyn Institute of Nuclear Physics, M.~V.~Lomonosov Moscow State University, 119991, Moscow, Russia\\
\llap{$^c$}
Faculty of Physics, University of Vienna, Boltzmanngasse 5, A-1090 Vienna, Austria\\ 
\llap{$^d$}Faculty of Physics, M. V. Lomonosov Moscow State University, 119991 Moscow, Russia\\
E-mail: \email{dmitri\_melikhov@gmx.de}, \email{anastasiia.kozachuk@cern.ch}, \email{Nikolai.Nikitine@cern.ch}
}

\abstract{We present our recent results \cite{kmn} for long-distance QCD effects in the flavour-changing neutral current 
radiative leptonic decays $B\to\gamma\ell^+\ell^-$, $\ell=\{e,\mu\}$. 
One encounters here two distinct types of long-distance effects: those encoded in the $B\to\gamma$ transition form factors induced by the 
$b\to q$ quark currents, 
and those related to the charm-loop effects. 

We calculate the $B\to\gamma$ form factors in a broad range of the momentum transfers 
making use of the relativisitc dispersion approach based on the constituent quark picture which has proven to provide reliable 
predictions for many weak-transition form factors. 

Concerning the description of the charm-loop contributions, we point out two observations:  
First, the precise description of the charmonium resonances, in particular, the relative phases 
between $\psi$ and $\psi'$, has impact on the differential distributions and on the forward-backward asymmetry, $A_{\rm FB}$, 
in a broad range of $q^2\ge 5$ GeV$^2$. Second, the shape of $A_{\rm FB}$ 
in $B\to\gamma\ell^+\ell^-$ and in $B\to V\ell^+\ell^-$ ($V$ the vector meson) {\it in the $q^2$-region between $\psi$ and $\psi'$} 
provides an unambiguous probe of the relative phases between $\psi$ and $\psi'$. 
Fixing the latter will lead to a strong reduction of the theoretical uncertainties in $A_{\rm FB}$ at $q^2=5-9$ GeV$^2$ where it 
has the sensitivity to physics beyond the SM.}

\FullConference{EPS-HEP 2017, European Physical Society conference on High Energy Physics\\
		5-12 July 2017\\
		Venice, Italy}

\begin{document}

\section{Motivation}
\label{sec-intro}
Rare $b\to q$ ($q=d,s$) transitions proceed through flavor changing neutral currents (FCNC) forbidden at tree level in the Standard Model. 
Because of possible contributions of new particles to the loops, the corresponding decays may be sensitive to physics beyond the Standard Model. 
Presently, a number of tensions between the predictions of the Standard Model and the experimental measurements in weak $B$-decays,
e.g. in $B\to K \ell^+\ell^-$ and $B\to K^* \ell^+\ell^-$ at the level of $2-3\sigma$ have been reported (see e.g. \cite{Diego} and references therein). 
If these deviations are confirmed at a larger statistics, other FCNC $B$-decays, such as rare radiative leptonic $B_{s,d}\to \gamma \ell^+\ell^-$ decays 
may provide important complementary information on deviations from the SM. Our analysis \cite{kmn} focuses on long-distance QCD effects in 
radiative leptonic $B$-decays which have a rich and a complicated structure. In this report we present some of the essential results from \cite{kmn}. 

\section{Theoretical basics}
The FCNC $b\to q$ transitions are described by the effective Hamiltonian \cite{Heff}
\begin{eqnarray}
\label{Heff}
H_{\rm eff}^{b\to q}=\frac{G_F}{\sqrt{2}}V^*_{tq}V_{tb}\sum_i C_i(\mu) O^{b\to q}_i(\mu),  
\end{eqnarray}
where $G_F$ is the Fermi constant, $C_i$ are the Wilson coefficients and $O_i$ are the basis operators of this expansion. 
The Wilson coefficients contain the short-distance effects describing the contributions of ``heavy'' degrees of 
freedom---$t$-quarks and $W$- and $Z$-bosons---in the loops. 
The operators $O_i$ involve the ``light'' degrees of freedom such as $b$, 
$c$ and the light quarks, the leptons, and the photon. 
Whereas the heavy degrees of freedom have been integrated out, the light degrees of freedom remain dynamical and propagate in the loops. 

One encounters two kinds of long-distance QCD effects in rare radiative $B$-decays: (i) those encoded in the $B\to\gamma$ transition form factors of 
the basis operators in (\ref{Heff}) and (ii) those related to the virtual $u$ and $c$ quarks in the loops. A serious challenge is related 
to the theoretical description of $c$-quark loops which lead to the appearance of the narrow charmonium resonances 
in the physical region of $B$-decays.  

\noindent(i) The transition form factors are defined as follows \cite{kruger,mn2004}\footnote{Our notations and conventions are: 
$\gamma^5=i\gamma^0\gamma^1\gamma^2\gamma^3$, 
$\sigma_{\mu\nu}=\frac{i}{2}[\gamma_{\mu},\gamma_{\nu}]$, 
$\varepsilon^{0123}=-1$, $\epsilon_{abcd}\equiv
\epsilon_{\alpha\beta\mu\nu}a^\alpha b^\beta c^\mu d^\nu$, 
$e=\sqrt{4\pi\alpha_{\rm em}}$.}:
\begin{eqnarray}
\label{real}
\label{ffs}
\nonumber
\langle
\gamma^* (k,\,\epsilon)|\bar s \gamma_\mu\gamma_5 b|\bar B_s(p)\rangle 
&=& 
i\, e\,\epsilon^*_{\alpha}\, \left ( g_{\mu\alpha} \, k'k-k'_\alpha k_\mu \right )\,\frac{F_A(k'^2,k^2)}{M_{B_s}}, 
\\
\langle \gamma^*(k,\,\epsilon)|\bar s\gamma_\mu b|\bar B_s(p)\rangle
&=& 
e\,\epsilon^*_{\alpha}\,\epsilon_{\mu\alpha k' k}\frac{F_V(k'^2,k^2)}{M_{B_s}},   
\\
\langle\gamma^*(k,\,\epsilon)|\bar s \sigma_{\mu\nu}\gamma_5 b|\bar B_s(p) 
\rangle\, k'^{\nu}
&=& 
e\,\epsilon^*_{\alpha}\,\left( g_{\mu\alpha}\,k'k- k'_{\alpha}k_{\mu}\right)\, 
F_{TA}(k'^2, k^2), 
\nonumber
\\
\langle
\gamma^*(k,\,\epsilon)|\bar s \sigma_{\mu\nu} b|\bar B_s(p)\rangle\, k'^{\nu}
&=& 
i\, e\,\epsilon^*_{\alpha}\epsilon_{\mu\alpha k' k}F_{TV}(k'^2, k^2).
\nonumber 
\end{eqnarray}
The form factors are functions of two variables, $k'^2$ and $k^2$, where the momentum $k'$ is emitted from the $b\to s$ weak transition current, whereas the 
momentum $k$ is emitted by one of the valence quarks of the $B$-meson. 

\begin{figure}[H]
\centering    
\includegraphics[width=3cm,clip]{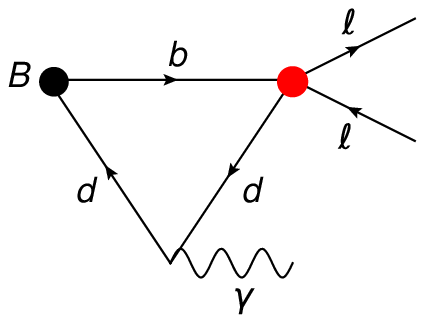}     
\hspace{8mm}
\includegraphics[width=3cm,clip]{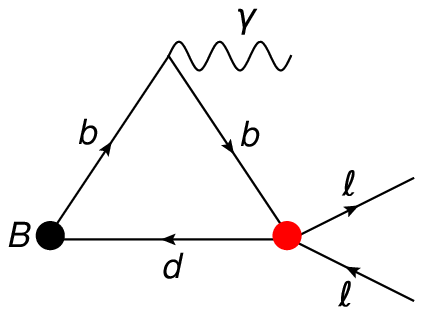}\\
\hspace{8mm}
\includegraphics[width=4cm,clip]{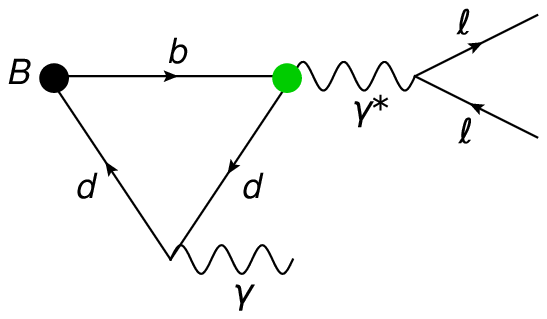}
\hspace{8mm}
\includegraphics[width=4cm,clip]{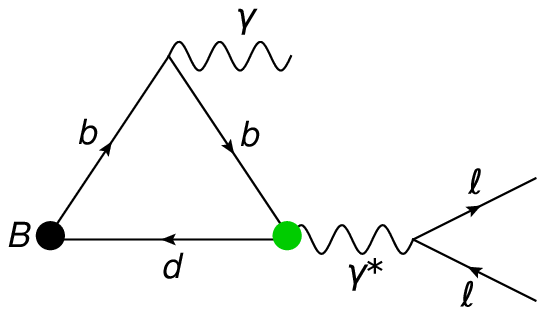}
\caption{Diagrams, corresponding to the case when the real photon is emitted 
by the valence quark of the $B$-meson and the lepton pair is coupled to the weak effective vertex. Red blobs denote the four-fermion operators and 
green blobs correspond to the electromagnetic penguin operator.}
\label{fig-effvertex} 
\end{figure}

There are two distinct types of form factors necessary for the description of the $B\to \gamma \ell^+\ell^-$. 
Fig.~\ref{fig-effvertex} shows diagrams describing amplitudes of the first type: the real photon is emitted by the 
valence quarks of the $B$-meson, whereas the virtual photon is emiited from the weak-transition vertex. 
These amplitudes correspond to $k^2=q^2$ and $k^2=0$. An essential feature of the form factors $F(q^2,0)$ is the absence of hadron singularites in the physical 
region of the radiative decay $0<q^2<(M_B-2m_l)^2$. The properties of the corresponding form factors have been discussed 
in \cite{kruger} where a realistic parametrization of these form factors have been proposed. This form-factor model has been applied to the description 
of $B\to \gamma \ell^+\ell^-$ decays in \cite{mn2004}. In \cite{kmn2016}, the form factors $F_{A,V}(q^2,0)$ have been studied in more detail, making use 
of the relativistic dispersion approach based on the constituent quark picture \cite{m}, which proved to provide a reliable description of a large class 
of weak-transition form factors \cite{ms}. The spectral representations for $F_{A,V}(q^2,0)$ in terms of the $B$-meson 
wave function were derived and reliable numerical estimates were obtained. Ref.~\cite{kmn} extends the analysis of \cite{kmn2016}: it calculates also 
the form factors $F_{TA,TV}(q^2,0)$ within the dispersion approach and provides the numerical results for these quantities. 

\begin{figure}[H]
\centering
\label{fig-vmd}     
\includegraphics[width=3cm,clip]{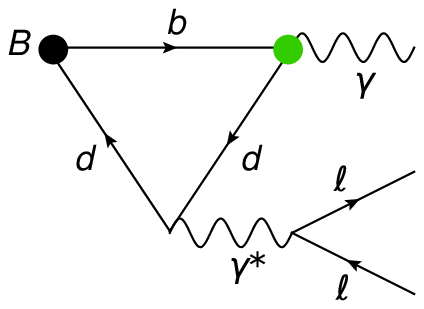}     
\hspace{10mm}
\includegraphics[width=3cm,clip]{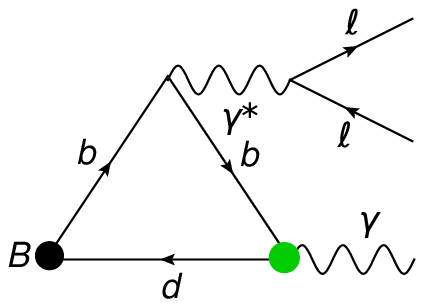}
\caption{Diagrams, corresponding to the case when the real photon is emitted from the effective weak vertex, 
while the virtual one is emitted from the valence quark of the $B$-meson}.
\end{figure}

Another type of contributions corresponds to the real-photon emission from the FCNC penguin, whereas the virtual photon is 
emitted by the valence quark (Fig.~\ref{fig-vmd}). These diagrams lead to the form factors $F_{TA,TV}(0,q^2)$. 
The diagram of Fig.~\ref{fig-vmd}(b) is easy as it does not contain hadron singularities in the physical decay region:  
the nearest singularity in the $q^2$ channel lies at $q^2=M_{\Upsilon}^2$, i.e. far beyond the physical region. 
On the contrary, the diagram of Fig.~\ref{fig-vmd}(a) involves the light vector mesons 
in the physical region. For the description of this amplitude, a gauge-invariant version of vector meson dominance is used. 
Further details of the calculation can be found in \cite{mn2004}.

\section{Charming loops}
A serious problem is posed by the contributions of four-quark operators from the effective Hamiltonian, which generate 
the $u$- and the $c$-quark loops. Crucial is the contribution of the charm loops as the narrow charmonium states appear
in the physical decay region. 

In the factorization approximation (i.e. neglecting gluon exchanges between the $B$-meson valence-quark loop and the 
$c$-quark loop attached to the photon) the contribtions of a vector charmonium state may be unambiguously calculated \cite{ali,sehgal}: 
the contribution of any vector resonance is positive-defined and is given via its decay constant $f_v$. 
However, as is well-known, the factorization approximation leads to the $B\to K^*\psi$ and $B\to K^*\psi'$ amplitudes strongly 
underestimating the known $B\to K^*(\psi,\psi')$ rates. 
Therefore, nonfactorizable gluons yield the dominant contribution, at least in the $q^2$ regions near $\psi$ and $\psi'$. 
In \cite{khodj}, the nonfactorizable gluon exchanges have been calculated at $q^2\ll 4m_c^2$; it was then 
concluded that the extrapolation to larger $q^2$ of the full amplitude---the sum of factorizable and nonfactorizable contributions---suggests 
an opposite sign of the $\psi$ and $\psi'$ contributions to the $B\to K^*\ell^+\ell^-$ amplitude, such that the full $\psi'$ contribution is negative. 
Because of the similarities between the amplitudes of the processes with the neutral vector meson and the photon in the final state, 
the same pattern of the $\psi$ and $\psi'$ contributions in $B\to \gamma \ell^+\ell^-$ would be a reasonable assumption. 

Since the conclusion about the opposite-sign phases of $\psi $ and $\psi'$ is based on the extrapolation, 
it would be useful to have also alternative checks of this statement. The LHCb collaboration studied the relative 
phase between the $\psi$ and $\psi'$ in the $B\to K\ell^+\ell^-$ decays and reported that the similar-phase and the opposite-phase cases 
cannot be discriminated using the available data \cite{lhcb}. 

We therefore consider two different possibilites: the same-phase and the opposite-phase 
$\psi$ and $\psi'$ contributions.\footnote{We emphasize 
that after nonfactrizable gluon exchanges are taken into account, the pattern of the $\psi$ and $\psi'$ 
contributions in $B\to V\ell^+\ell^-$ and in $B\to P\ell^+\ell^-$ decays may well differ from each other.} 
We will show that the behaviour of the forward-backward asymmetry 
$A_{FB}$ in $B\to \gamma \ell^+\ell^-$ and in $B\to V \ell^+\ell^-$ decays in the $q^2$-range between $\psi$ and $\psi'$ 
provides an unambiguos probe of the relative phases of these vector charmonia contributions. 

\begin{figure}[b]
\centering
\includegraphics[width=6cm]{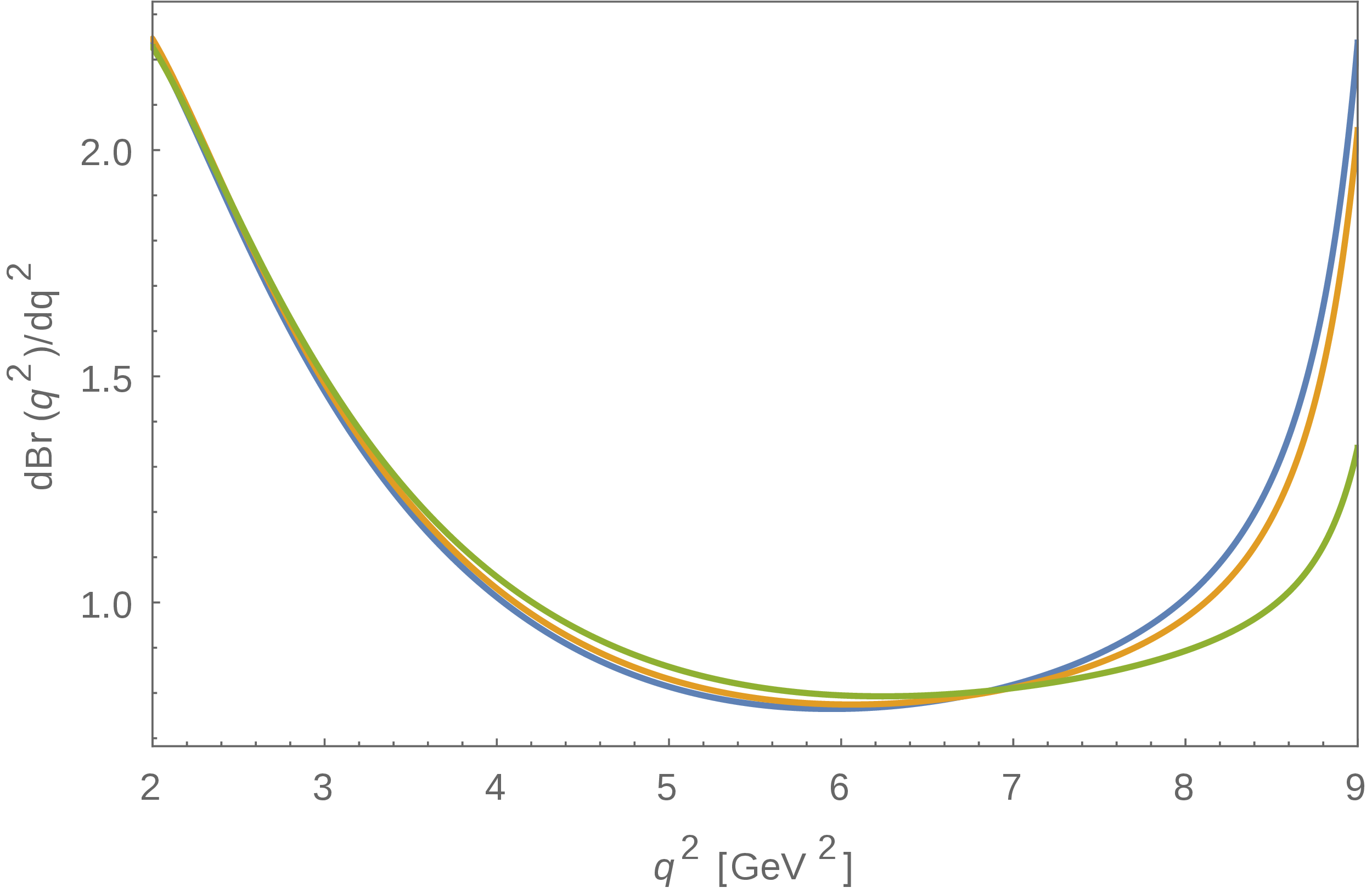}
\hspace{10mm}
\includegraphics[width=6cm]{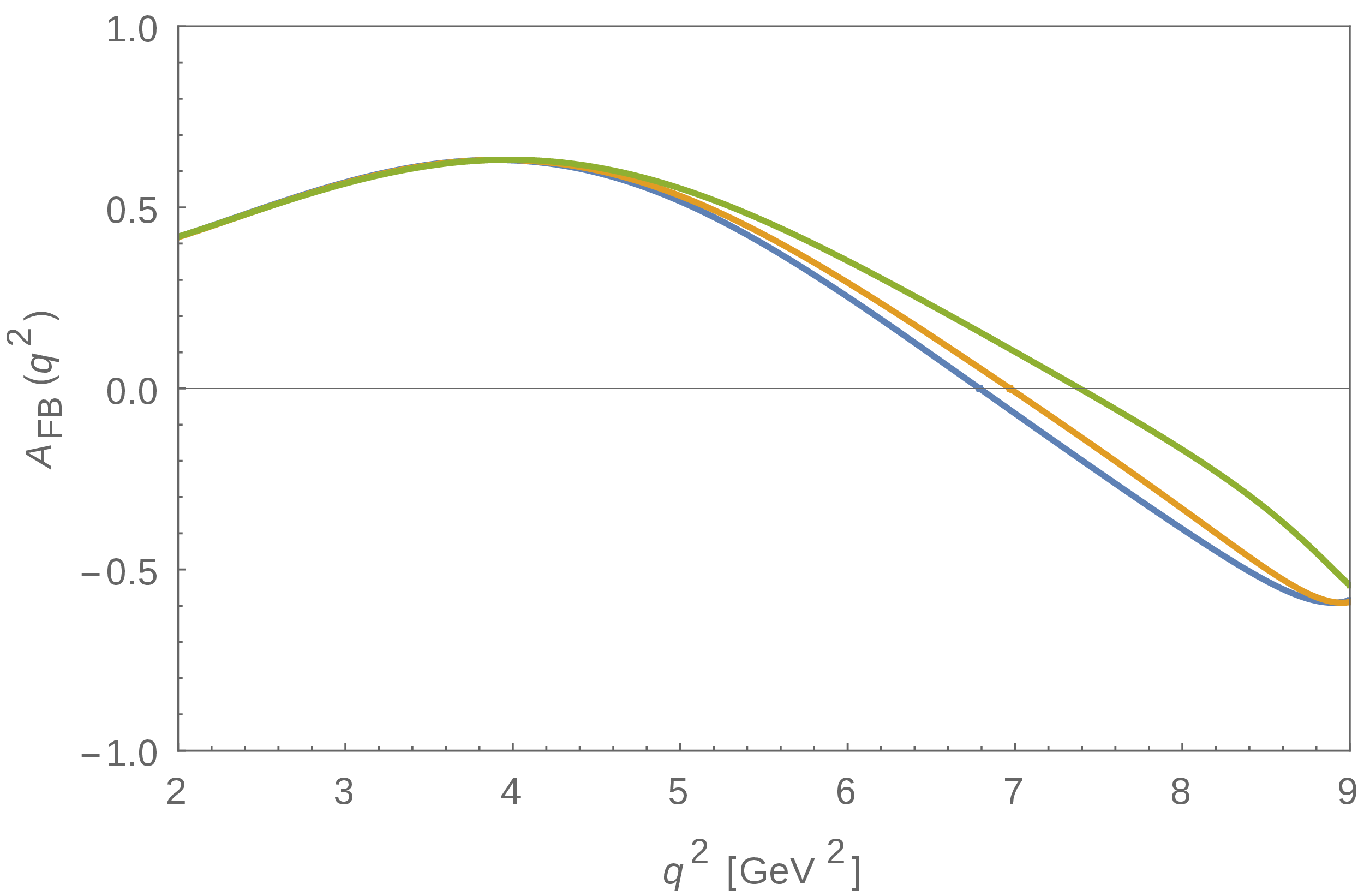}
\caption{The differential branching ratio $d\Gamma/dq^2$ and the forward-backward asymmetry $A_{FB}(q^2)$ for $\bar{B}_{s}\to\gamma\mu^+\mu^-$ 
in the range $q^2=[2,9]$ GeV$^2$. Blue line - the description of $\psi$ and $\psi'$ from \cite{ali}; 
orange line - the description from \cite{ali} with the $\psi'$ phase reversed; 
green line - the description of $\psi$ and $\psi'$ from \cite{khodj}.}
\label{fig-br}      
\end{figure}

\section{Numerical Results}
\label{sec-nresults}
We now present our numerical results for $\bar{B}_{s}\to \gamma \ell^+\ell^-$. Fig.~\ref{fig-br} shows the differential distributions in 
$\bar{B}_{s}\to \gamma\mu^+\mu^-$ decays for three different ways of describing the $\psi$ and $\psi'$ contributions. 
In the region $q^2=5-9$ GeV$^2$, where the asymmetry has a zero point, the distributions are sensitive to the details of the 
description of charmonia contributions.

\begin{figure}[h!]
\centering
\includegraphics[width=9cm]{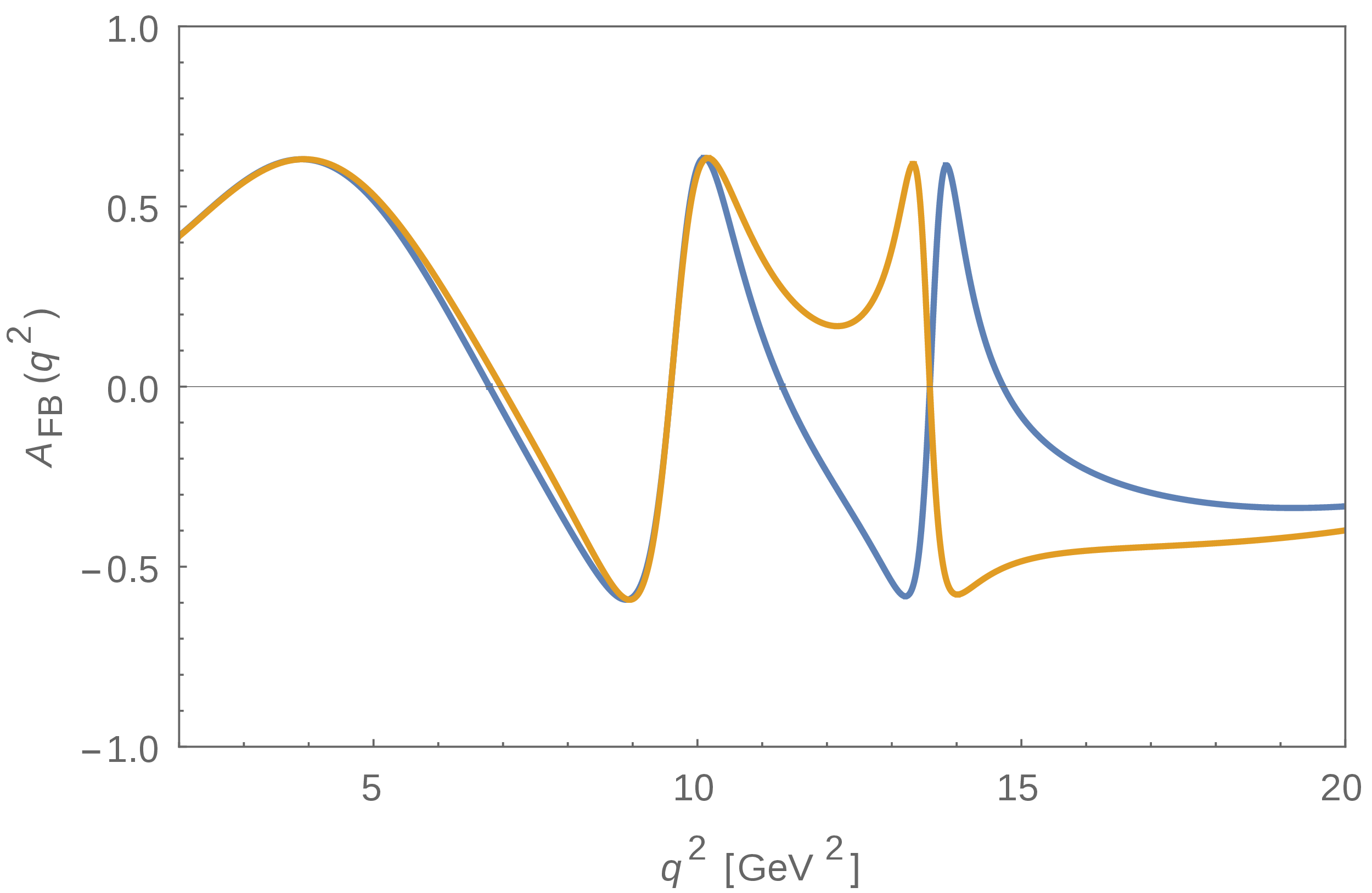}
\caption{Forward-backward asymmetries for $\bar{B}_{s}\to\gamma\mu^+\mu^-$. Blue line - the description of $\psi$ and $\psi'$ from \cite{ali}; 
orange line - the description from \cite{ali} with the $\psi'$ phase reversed.}
\label{fig-ratios}      
\end{figure}

Fig.~\ref{fig-ratios} gives $A_{FB}$ in a broad range of $q^2$, including the charmonia region. Obviously, the shape of $A_{FB}$ 
in the region between  $\psi$ and $\psi'$ is particularly sensitive to the relative sign between these states. 

Table~\ref{tab-ratios} summarizes our numerical predictions for the branching ratios. We present the branching ratios integrated over 
two different regions: over the range $q^2\in[1;6]\,\mathrm{GeV^2}$, which does not include any resonance, and over the whole range of the 
invariant mass of the lepton pair. In the latter case we include the contributions of the light vector mesons and the 
vector charmonia using a simple Breit-Wigner formula \cite{ali}.  

\begin{table}[H]
\centering
\caption{Our results for the branching ratios of $\bar{B}_{d,s}\to l^+l^-\gamma$ decays for different ranges of $q^2$: 
for $q^2\in[1;6]\,\mathrm{GeV^2}$ and for the full range. The value of the photon energy cut is $500\,\mathrm{MeV}$ }
\label{tab-ratios}
\vspace{.2cm}

\begin{tabular}{l|ll}
\hline
$q^2$-range                    & $[1;6]\,\mathrm{GeV^2}$ & Full \\
\hline
\protect\( Br\left ( B\to e^+e^-\gamma\right )\,\times\, 10^{10}\protect\) &

                           $0.12$    &
                           $4.6$ \\
\protect\( Br\left ( B\to \mu^+\mu^-\gamma\right )\,\times\, 10^{10}\protect\) &
                           $0.12$    &
                           $1.5$ \\
\protect\( Br\left ( B_s\to e^+e^-\gamma\right )\,\times\, 10^{9}\protect\) &
                           $0.76$    &
                           $18.4$ \\
\protect\( Br\left ( B_s\to \mu^+\mu^-\gamma\right )\,\times\, 10^{9}\protect\) &
                           $0.73$     &
                           $11.6$ \\
\hline
\end{tabular}
\end{table}
\section{Conclusion}
We presented the outcomes of our resent study of $B\to\gamma \ell^+\ell^-$ decays in the SM. The emphasis of our analysis was laid on the long-range QCD contributions to the 
$B\to\gamma \ell^+\ell^-$ amplitude. It should be emphasized that radiative leptonic decays have a more complicated structure of long-distance effects compared 
to rare FCNC semileptonic decays. 

Our main findings may be summarized as follows: 

\vspace{.2cm}
\noindent
1. We obtained predictions for the $B\to \gamma$ tranition form factors within the relativistic dispersion approach based on the constituent quark picture.
Our numerical estimates correspond to the parameters of the meson wave functions updated according to the latest values of the decay constants of 
pseudoscalar and vector mesons. Comparison with the available results from other approaches suggest the accuracy of our predictions 
at the level of about 10\%.  

\vspace{.2cm}
\noindent
2. We studied in detail long-distance effects related to charm loops. Making use of different prescriptions of the $\psi$ and $\psi'$ phases, we observed a 
sizeable sensitivity of the differential distributions and the forward-backward asymmetry in the region $q^2=5-9$ GeV$^2$ to the sign of the $\psi'$ 
contribution to the amplitude. We point out that the experimental measurement of the shape of $A_{FB}$ in $B\to V\ell^+\ell^-$ or in 
$B\to \gamma \ell^+\ell^-$ decays 
in the $q^2$-region between $\psi$ and $\psi'$ provides a clean probe of the relative $\psi$ and $\psi'$ phases. As soon as the latter are fixed, 
the forward-backward asymmetry $A_{FB}$ in the region of $q^2=5-9$ GeV$^2$, where it has a sensitiity to the extentions of the SM, may be reliably 
calculated and used for the search of the new physics effects. 

\section*{Acknowledgments}
\noindent
The authors are grateful to Diego Guadagnoli for numerous useful discussions. 
D.~M. acknowledges support from the Austrian Science Fund (FWF) under project P29028; 
A.~K. and N.~N. were supported by the Russian Science Fund under grant 16-12-10280. 

\end{document}